\documentclass[10pt,american]{IEEEtran}
\usepackage[T1]{fontenc}
\usepackage[latin9]{inputenc}
\usepackage{amsmath}
\usepackage{amssymb}
\usepackage{graphicx}
\usepackage{esint}

\makeatletter

\usepackage{graphics}
\usepackage{cite}
\usepackage{multirow}

\usepackage{babel}

\makeatother

\usepackage{babel}
\begin{document}

\title{Reconstruction of Complex Baseband Signals via $M$-Periodic Nonuniform
Bandpass Sampling and Least-Squares Optimal Time-Varying FIR Filters}
\author{\selectlanguage{english}%
\IEEEauthorblockN{Håkan Johansson\\
 } \IEEEauthorblockA{Dept. Electrical Engineering\\
Linköping University, Sweden\\
Email: hakan.johansson@liu.se}}
\maketitle
\selectlanguage{american}%
\begin{abstract}
This paper considers the reconstruction of digital complex baseband
signals from $M$-periodically nonuniformly sampled real bandpass
signals. With such a sampling, bandpass signals with arbitrary frequency
locations can be sampled and reconstructed, as opposed to uniform
sampling which requires the signal to be within one of the Nyquist
bands. It is shown how the reconstruction can be carried out via an
$M$-periodic time-varying finite-length impulse response (FIR) filter or, equivalently, a set of $M$
time-invariant FIR filters. Then, a least-squares design method is
proposed in which the $M$ filter impulse responses are computed in
closed form. This offers minimal filter orders for a given desired
bandwidth. This is an advantage over an existing technique where ideal
filters are first derived (ensuring perfect reconstruction) and then
windowed and truncated, which leads to suboptimal filters and thus
higher filter orders and implementation complexity. A design example
illustrates the efficiency of the proposed design technique.
\end{abstract}

\section{Introduction}

\label{section:introduction} \label{Introduction}

Traditional communication receivers use analog mixers (modulators)
and filters prior to sampling \cite{Mirabbasi_2000,Venosa_2012}.
The sampling is either carried out for a real low-IF signal using
one analog-to-digital converter (ADC), or for a complex zero-IF signal using
two ADCs (at half the low-IF sampling rate) for the real (I) and imaginary
(Q) parts, respectively. In the zero-IF case, the desired digital
complex baseband signal (also referred to as complex envelop) is obtained
directly after the sampling. In the low-IF case, additional digital
modulation and filtering are required. An alternative is to sample
the real bandpass signal directly and then digitally reconstruct the
complex baseband signal. Thereby, analog mixing (demodulation) is
no longer needed. When the bandpass signal is located within one of
the Nyquist bands (for a given sampling frequency), uniform sampling
can be used as aliasing is then avoided and the complex baseband signal
can be readily obtained through regular digital filtering and modulation.
This causes a challenge for flexible radio receivers as different
sampling frequencies then have to be used for signals in different
frequency bands, even if they have the same bandwidth. A solution
to this problem is to use $M$-periodic nonuniform sampling with a
fixed sampling rate, as considered recently in \cite{Wahab_2022}.
In that case, bandpass signals with arbitrary frequency locations
can be handled. In practice, such sampling can be implemented using
a nonuniform-sampling $M$-channel time-interleaved ADC (TI-ADC),
as opposed to conventional uniform-sampling TI-ADCs \cite{Black_80,Johansson_2024_bookchapter}.
A challenge here is to design and implement reconstruction filters
with low implementation complexity, which is addressed in this paper.
Before proceeding, it is noted that reconstruction techniques for
$M$-periodic sampling have been proposed in several papers in the
past. However, as discussed in \cite{Wahab_2022}, they are either
not applicable or have limitations for the reconstruction problem
considered here. The contribution of the paper is as follows.

The reconstruction
is carried out via an $M$-periodic time-varying FIR filter, which can
be described by a set of $M$ time-invariant FIR filters.
Based on that, it is shown how the whole sampling and reconstruction
system can be described by an $M$-periodic time-varying frequency
function or, equivalently, a set of $M$ time-invariant frequency
functions. This is different from \cite{Wahab_2022} in which the
sampling and reconstruction were described in terms of downsamplers,
upsamplers, and filters (which do not coincide with the filters mentioned
above), which correspond to a multirate filter-bank representation,
having a distortion function (ideally one) and aliasing functions
(ideally zero). Both alternative representations are always feasible
for $M$-periodic time-varying systems \cite{Vaidyanathan_93,Mehr_02},
but an advantage of the time-varying system representation in this
paper lies in the design. Specifically, it will be shown how the $M$
impulse responses\footnote{For even $M$, it suffices to design and implement $M/2$ filters,
as explained in Section \ref{sec:Efficient-Implementation}. } can be determined separately in closed form, via least-squares minimization
of error functions that incorporate the corresponding frequency functions.
This gives minimal filters orders for a given desired bandwidth. In
\cite{Wahab_2022}, ideal filters were first derived and then windowed
and truncated which results in suboptimal filters due to the windowing
technique \cite{Jackson_96,Wanhammar_11}. In addition, in \cite{Wahab_2022},
closed-form solutions were derived only for $M=2,3,4$, and the derivations
become more difficult as $M$ increases. The technique in this paper,
on the other hand, works equally well for arbitrary $M$. In addition,
the derivations in this paper, which differ from those in \cite{Wahab_2022},
offer further insights into the reconstruction problem.

Following this introduction, Section \ref{sec:Sampling-and-Reconstruction}
considers the sampling and reconstruction problem and the proposed
reconstruction. Sections \ref{sec:Efficient-Implementation} and \ref{sec:Least-Squares-Design}
consider efficient implementation and least-squares design, respectively.
Section \ref{sec:Example} provides design examples whereas Section
\ref{sec:Conclusion} concludes the paper.

\section{Sampling and Reconstruction\label{sec:Sampling-and-Reconstruction}}

\begin{figure}[t!]
\centering \scalebox{0.8}{\includegraphics[scale=0.9]{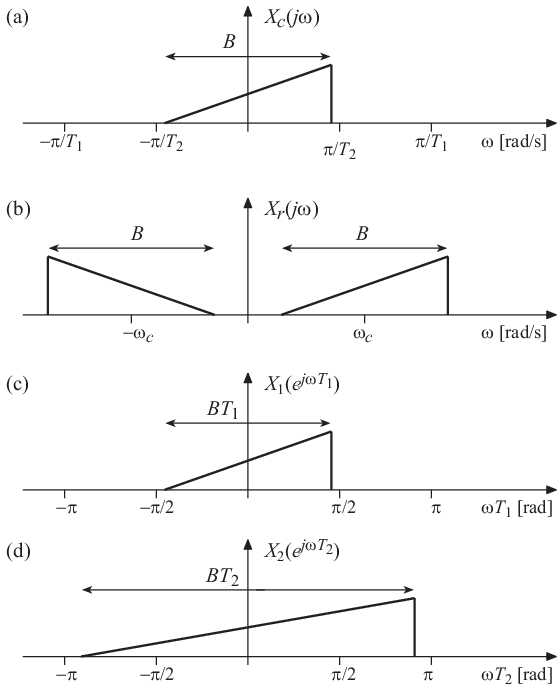}}
\caption{Example spectra.}

\label{fig:principle_spectra}
\end{figure}

The point of departure is a complex continuous-time baseband signal
$x_{c}(t)$ with bandwidth $B$ (in practice obtained from a digitally
generated baseband signal via digital-to-analog conversion). The corresponding
real continuous-time bandpass signal, $x_{r}(t)$, centered at $\omega_{c}$,
is obtained from $x_{c}(t)$ as
\begin{equation}
x_{r}(t)=\Re\{x_{c}(t)e^{j\omega_{c}t}\}=\frac{1}{2}\left(x_{c}(t)e^{j\omega_{c}t}+x_{c}^{*}(t)e^{-j\omega_{c}t}\right),
\end{equation}
where $*$ denotes conjugate. The corresponding Fourier transform
is
\begin{equation}
X_{r}(j\omega)=\frac{1}{2}\left(X_{c}(j\omega-j\omega_{c})+X_{c}^{*}(-j\omega-j\omega_{c})\right).\label{eq:FT_xr}
\end{equation}
Example spectra for the two signals are seen in Fig. \ref{fig:principle_spectra}(a)
and (b). Further, let $x_{1}(n)$ and $x_{2}(m)$ correspond to sampled
versions of $x_{c}(t)$ with sampling periods $T_{1}$ and $T_{2}$,
respectively. That is,
\begin{equation}
x_{1}(n)=x_{c}(nT_{1})\quad x_{2}(m)=x_{c}(mT_{2}).
\end{equation}
The sampling frequencies are thus $f_{s1}=1/T_{1}$ and $f_{s2}=1/T_{2}$.
To enable sampling of $x_{r}(t)$ and further signal reconstruction,
we require $f_{s1}>2B$. Further, throughout this paper, we assume
that $f_{s1}=2f_{s2}$ because $x_{2}(m)$ is then readily obtained
from $x_{1}(n)$ via downsampling by two, i.e.,
\begin{equation}
x_{2}(m)=x_{1}(2m).
\end{equation}
Under the above assumptions, we have the Fourier-domain relations
\begin{equation}
X_{k}(e^{j\omega T_{k}})=f_{sk}X_{c}(j\omega),\;|\omega|<\pi B/2,\;k=1,2,\label{eq:FT_relation}
\end{equation}
and
\begin{equation}
X_{2}(e^{j\omega T_{2}})=\frac{1}{2}X_{1}(e^{j\omega T_{1}}),\;|\omega T_{2}|<\pi.
\end{equation}
Example spectra of these discrete-time signals, corresponding to the
continuous-time signals in Fig. \ref{fig:principle_spectra}(a) and
(b), are as seen in Fig. \ref{fig:principle_spectra}(c) and (d).

\subsection{Nonuniform Sampling and Proposed Reconstruction}

\begin{figure}[t!]
\centering \scalebox{0.8}{\includegraphics[scale=0.9]{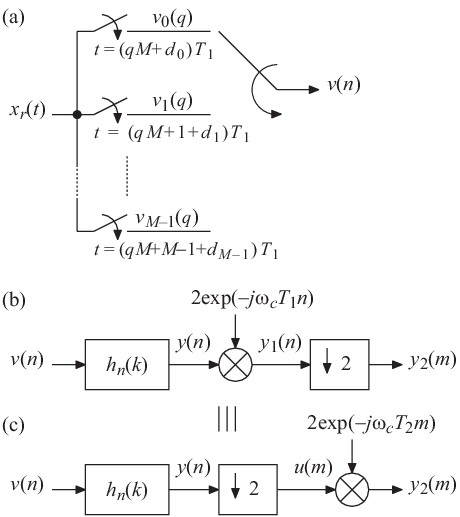}}
\caption{(a) Generation of $v(n)$ using an $M$-channel TI-ADC. (b) Proposed
reconstruction scheme using an $M$-periodic time-varying filter $h_{n}(k)$.
(c) Equivalent scheme with the modulation after downsampling. The
scheme can be efficiently implemented as outlined in Section \ref{sec:Efficient-Implementation}.}

\label{fig:TI-ADC and reconstr}
\end{figure}

The objective is to first sample the real signal $x_{r}(t)$ and then,
from the obtained samples, generate (approximate) the digital complex baseband
signal $x_{2}(m)$. Here, $M$-periodic nonuniform sampling of $x_{r}(t)$
is adopted (as in \cite{Wahab_2022}), with the sampling period $T_{1}$
to fulfill the sampling theorem. Hence, the sampling generates the
sequence $v(n)$ according to
\begin{equation}
v(n)=x_{r}(nT_{1}+d_{n}T_{1}),\quad d_{n}=d_{n+M},
\end{equation}
where $d_{n}$ represent time skews, i.e., $d_{n}T_{1}$ represent
deviations from a uniform-sampling grid $nT_{1}$. In practice, as
depicted in Fig. \ref{fig:TI-ADC and reconstr}(a), $v(n)$ can be
generated via an $M$-channel TI-ADC where each channel ADC operates
at the $M$ times lower sampling frequency $f_{s1}/M$. When the signal
$x_{r}(t)$ is located within one of the Nyquist bands ($|\omega|\in[pf_{s1}/2,(p+1)f_{s1}/2],p\in\{\mathbb{Z}\}_{\geq0}$),
uniform sampling can be used (i.e, all $d_{n}=0$) as aliasing is
then not introduced and reconstruction is not needed. However, for
arbitrary frequency locations, uniform sampling introduces aliasing
and the desired signal $x_{2}(m)$ cannot be generated from the uniform-grid
samples $x_{r}(nT_{1})$. In such cases, nonuniform sampling must
be adopted to enable the generation of $x_{2}(m)$ from the nonuniform-grid
samples $x_{r}(nT_{1}+d_{n}T_{1}$), in which case at least one of
$d_{n}$ is nonzero. The feasible selections of $d_{n}$ were considered
in \cite{Wahab_2022} and will not be further discussed here due to
the limited space.

Given $v(n)$, the objective is now to generate a new sequence $y_{2}(m)$
that ideally should equal (in practice approximate) the digital complex
baseband signal $x_{2}(m)$. To ease the derivation of the final implementation,
as well as the design, we will first generate $y_{1}(n)$ which ideally
should equal $x_{1}(n)$. Then, $y_{2}(m)$ is obtained via dowsampling
by two, i.e., $y_{2}(m)=y_{1}(2m).$ The sequence $y_{1}(n)$ is generated
by applying $v(n)$ as an input to an $M$-periodic linear system
(filter) characterized by the $M$-periodic impulse response $h_{n}(k)=h_{n+M}(k)$.
It is assumed that $h_{n}(k)$ is a finite-length impulse response
(FIR) filter of even order $N$, in which case the output $y_{1}(n)$
can be computed via linear convolution as\footnote{For simplicity in the mathematical expressions and design, even-order
non-causal filters are assumed. A causal filter is obtained by right-shifting
the noncausal filter $N/2$ samples. Odd-order filters can also be
handled after minor appropriate modifications.}
\begin{equation}
y(n)=\sum_{k=-N/2}^{N/2}v(n-k)h_{n}(k).\label{eq:y(n)}
\end{equation}
To see how to determine $h_{n}(k)$, we first rewrite $v(n)$ in terms
of its inverse Fourier transform, also utilizing \eqref{eq:FT_xr}
and the bandlimitation assumption, as 
\begin{eqnarray}
v(n) & = & x_{r}(nT_{1}+d_{n}T_{1})\nonumber \\
 &  & =\frac{1}{2\pi}\int_{|\omega|\in[\omega_{c}-B/2,\omega_{c}+B/2]}X_{r}(j\omega)e^{j\omega T_{1}(n+d_{n})}d\omega\nonumber \\
 & = & \frac{1}{4\pi}\int_{\omega_{c}-B/2}^{\omega_{c}+B/2}X_{c}(j\omega-j\omega_{c})e^{j\omega T_{1}(n+d_{n})}d\omega\nonumber \\
 &  & +\frac{1}{4\pi}\int_{-\omega_{c}-B/2}^{-\omega_{c}+B/2}X_{c}^{*}(-j\omega-j\omega_{c})e^{j\omega T_{1}(n+d_{n})}d\omega.\nonumber \\
\label{eq:v(n)}
\end{eqnarray}
Then, inserting \eqref{eq:v(n)} in \eqref{eq:y(n)}, one obtains
\begin{eqnarray}
y(n) & = & \frac{1}{4\pi}\int_{\omega_{c}-B/2}^{\omega_{c}+B/2}A_{n}(j\omega T_{1})X_{c}(j\omega-j\omega_{c})e^{j\omega T_{1}n}d\omega\nonumber \\
 &  & +\frac{1}{4\pi}\int_{-\omega_{c}-B/2}^{-\omega_{c}+B/2}\big(A_{n}(j\omega T_{1})X_{c}^{*}(-j\omega-j\omega_{c})\nonumber \\
 &  & \times e^{j\omega T_{1}n}\big)d\omega,
\end{eqnarray}
where
\begin{equation}
A_{n}(j\omega T_{1})=\sum_{k=-N/2}^{N/2}h_{n}(k)e^{-j\omega T_{1}(k-d_{n-k})}.
\end{equation}
Via the variable substitutions $\omega\rightarrow\omega T_{1}$, $\omega\rightarrow\omega-\omega_{c}$,
and $\omega\rightarrow\omega+\omega_{c}$, and utilizing \eqref{eq:FT_relation},
one finally obtains
\begin{eqnarray}
y(n) & = & e^{j\omega_{c}T_{1}n}\frac{1}{4\pi}\int_{-(B/2)T_{1}}^{(B/2)T_{1}}\big(A_{n}(j\omega T_{1}+j\omega_{c}T_{1})\nonumber \\
 &  & \times X_{1}(e^{j\omega T_{1}})e^{j\omega Tn}d(\omega T_{1})\big)\nonumber \\
 &  & +e^{-j\omega_{c}T_{1}n}\frac{1}{4\pi}\int_{-(B/2)T_{1}}^{(B/2)T_{1}}\big(A_{n}(j\omega T_{1}-j\omega_{c}T_{1})\nonumber \\
 &  & \times X_{1}^{*}(e^{j\omega T_{1}})e^{j\omega Tn}d(\omega T_{1})\big).\label{eq:y(n)_expression}
\end{eqnarray}
Utilizing the inverse Fourier transform in \eqref{eq:y(n)_expression},
it is seen that, if
\begin{equation}
A_{n}(j\omega T_{1})=\begin{cases}
1, & \omega T_{1}\in[\omega_{c}T_{1}-BT_{1}/2,\omega_{c}T_{1}+BT_{1}/2]\\
0, & \omega T_{1}\in[-\omega_{c}T_{1}-BT_{1}/2,-\omega_{c}T_{1}+BT_{1}/2],
\end{cases}\label{eq:An_ideal}
\end{equation}
then $y(n)=(1/2)\times x_{1}(n)e^{j\omega_{c}T_{1}n}$. Hence, we
obtain the sequence $y_{1}(n)$ via digital modulation as $y_{1}(n)=2y(n)e^{-j\omega_{c}T_{1}n}$.
Finally, $y_{2}(m)$ is obtained from $y_{1}(n)$ via downsampling
by two. The reconstruction scheme is depicted in Fig. \ref{fig:TI-ADC and reconstr}(b).
Moving the modulation to after the downsampling by two results in
the equivalent scheme in Fig. \ref{fig:TI-ADC and reconstr}(c).

\section{Efficient Implementation\label{sec:Efficient-Implementation}}

\begin{figure}[t!]
\centering \scalebox{0.8}{\includegraphics[scale=0.9]{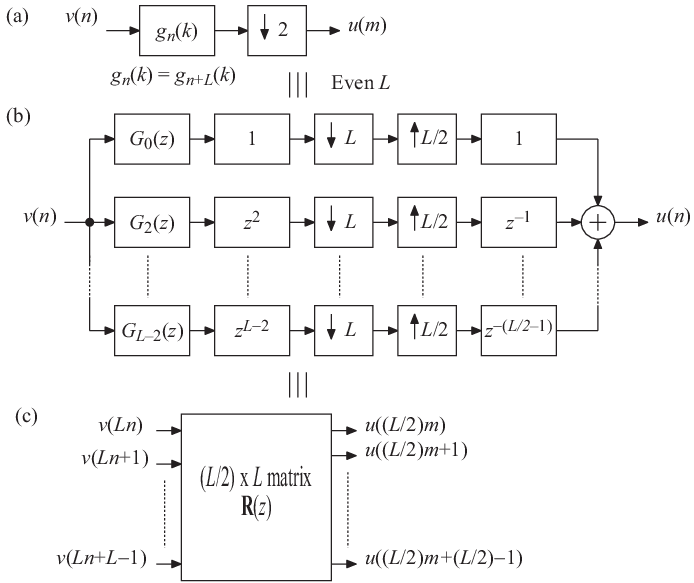}}
\caption{(a) Implementation of an $L$-periodic time-varying filter $g_{n}(k)$
followed by downsampling by two. (b) Equivalent scheme for even $L$,
using the corresponding set of $L/2$ time-invariant filters with
transfer functions $G_{n}(z),n=0,2,\ldots,L-2$. (c) Equivalent scheme
where all filtering operations run at the lower sampling rate.}

\label{fig:FB_equivalence}
\end{figure}

The scheme in Fig. \ref{fig:TI-ADC and reconstr} shows the principle
of the proposed reconstruction, where the periodic filter $h_{n}(k)$
operates at the high sampling rate $f_{s1}$. This representation is
convenient for the analysis and design, to be considered in Section
\ref{sec:Least-Squares-Design}, but a corresponding straightforward
implementation is inefficient. This section provides efficient implementations
via polyphase decomposition \cite{Vaidyanathan_93}, in which case
redundant operations are removed and all filtering operations are
carried out at the lower sampling rate $f_{s1}/M$. They are derived
via the equivalences seen in Fig. \ref{fig:FB_equivalence}. The scheme
in (b) follows from that in (a) by utilizing properties of upsamplers
and downsamplers \cite{Vaidyanathan_93,Johansson_06}. The scheme
in (c) follows from that in (b) via polyphase decomposition of the
filters $z^{n}G_{n}(z)$ according to

\begin{equation}
z^{n}G_{n}(z)=\sum_{l=0}^{L-1}z^{l}P_{nl}(z^{L}),
\end{equation}
 with $P_{nl}(z)$ denoting the $l$th polyphase component of $z^{n}G_{n}(z)$,
by which one can write
\begin{equation}
\left[\begin{array}{c}
G_{0}(z)\\
z^{2}G_{2}(z)\\
\vdots\\
z^{L-2}G_{L-2}(z)
\end{array}\right]={\mathbf{R}}(z^{L})\left[\begin{array}{c}
1\\
z\\
\vdots\\
z^{L-1}
\end{array}\right]\label{eq:54}
\end{equation}
 where ${\mathbf{R}}(z)$ is the $(L/2)\times L$ polyphase matrix
\begin{equation}
{\mathbf{R}}(z)=\left[\begin{array}{cccc}
P_{00}(z) & P_{01}(z) & \cdots & P_{0,L-1}(z)\\
P_{20}(z) & P_{21}(z) & \cdots & P_{2,L-1}(z)\\
\vdots & \vdots & \ddots & \vdots\\
P_{L-2,0}(z) & P_{L-2,1}(z) & \cdots & P_{L-2,L-1}(z)
\end{array}\right].\label{eq:55}
\end{equation}
There are now two different cases to consider, depending on whether
$M$ is even or odd.

\subsection{Even $M$}

Generating $y_{2}(m)$ from $y_{1}(n)$ via downsampling by two means
that the outputs of the odd-indexed filters $h_{2n+1}(k)$ will be
discarded when $M$ is even. Hence, $h_{2n+1}(k)$ will not be used,
which means that the reconstruction can be implemented with the scheme
in Fig. \ref{fig:FB_equivalence} with $L=M$ and $g_{2n}(k)=h_{2n}(k)$,
$n=0,2,\ldots,M-$2. As seen in Fig. \ref{fig:FB_equivalence}(c),
there are $M$ inputs to the matrix $R(z)$ which corresponds to a
block of $M$ samples from the TI-ADC output seen in Fig. \ref{fig:TI-ADC and reconstr}(a).
Further, in the design, it suffices to determine only $M/2$ impulse
responses.

\subsection{Odd $M$}

When $M$ is odd, all $M$ filters $h_{n}(k)$ will be used but, due
to the downsampling by two, in the order $h_{0}(k),h_{2}(k),\ldots h_{M-1}(k),h_{1}(k),h_{3}(k),\ldots h_{M-2}(k)$.
In this case, the reconstruction can be implemented with the scheme
in Fig. \ref{fig:FB_equivalence} with $L=2M$, $g_{2p}(k)=h_{2p}(k)$
for $p=0,1,\ldots,(M-1)/2$, and $g_{2p+M-1}(k)=h_{2p-1}(k)$ for
$p=1,2,\ldots,(M-1)/2$. Further, as seen in Fig. \ref{fig:FB_equivalence}(c),
there are in this case $L=2M$ inputs to the matrix $R(z)$ which
correspond to two consecutive blocks of $M$ samples from the TI-ADC
output in Fig. \ref{fig:TI-ADC and reconstr}(a).

\section{Least-Squares Design\label{sec:Least-Squares-Design}}

This section proposes a least-squares design technique for the $M$
complex-valued (since $A_{n}(j\omega T_{1})$ in \eqref{eq:An_ideal}
are not conjugate symmetric) impulse responses $h_{n}(k)$, $n=0,1,\ldots,M-1$.
In this proposal, $h_{n}(k)$ are determined separately and in closed
form via matrix inversion. To this end, it follows from \eqref{eq:y(n)_expression}
and \eqref{eq:An_ideal} that it is appropriate to use the $M$ error
functions
\begin{eqnarray}
P_{n} & = & \frac{1}{2\pi}\int_{\omega_{1}T}^{\omega_{2}T}|A_{n}(j\omega T_{1})-1|^{2}d(\omega T_{1})\nonumber \\
 &  & +\frac{1}{2\pi}\int_{-\omega_{2}T}^{-\omega_{1}T}|A_{n}(j\omega T_{1})|^{2}d(\omega T_{1}),\label{eq:Power functions}
\end{eqnarray}
where $\omega_{1}T_{1}=\omega_{c}T_{1}-BT_{1}/2$ and $\omega_{2}T_{1}=\omega_{c}T_{1}+BT_{1}/2$.
With
\begin{equation}
{\mathbf{h}}_{n}=\left[\begin{array}{cccc}
h_{n}(-N/2) & h_{n}(-N/2+1) & \ldots & h_{n}(N/2)\end{array}\right]^{T},\label{eq:h_vectors}
\end{equation}
where $T$ denotes transpose, each ${\mathbf{h}}_{n}$ is then obtained
by minimizing the corresponding error function $P_{n}.$ After some
algebraic manipulations and computations of integrals, one finds the
$M$ separate closed-form solutions
\begin{equation}
{\mathbf{h}}_{n}={\mathbf{S}}_{n}^{-1}{\mathbf{c}}_{n},\label{eq:h_optimal_solution}
\end{equation}
where ${\mathbf{c}}_{n}$ are $(N+1)\times1$ column vectors with
entries $c_{n,k},\ k=-N/2,-N/2+1,\ldots,N/2$, and ${\mathbf{S}}_{n}$
are $(N+1)\times(N+1)$ matrices with entries $s_{n,kp},\ k,p=-N/2,-N/2+1,\ldots,N/2$,
given by
\begin{equation}
c_{n,k}=\left\{ \begin{array}{lll}
(\omega_{2}T_{1}-\omega_{1}T_{1})/(2\pi), & k-d_{n-k}=0\\
\frac{\exp[j\omega_{2}T_{1}(k-d_{n-k})]}{j2\pi(k-d_{n-k})}\\
-\frac{\exp[j\omega_{1}T_{1}(k-d_{n-k})]}{j2\pi(k-d_{n-k})}, & k-d_{n-k}\neq0
\end{array}\right.\label{eq:c_n,k}
\end{equation}
\begin{equation}
s_{n,kp}=\left\{ \begin{array}{lll}
(\omega_{2}T_{1}-\omega_{1}T_{1})/\pi, & k=p\\
\frac{\sin[\omega_{2}T_{1}(k+d_{n-p}-p-d_{n-k})]}{\pi(k+d_{n-p}-p-d_{n-k})}\\
-\frac{\sin[\omega_{1}T_{1}(k+d_{n-p}-p-d_{n-k})]}{\pi(k+d_{n-p}-p-d_{n-k})}, & k\neq p.
\end{array}\right.\label{eq:s_n,kp}
\end{equation}

\section{Design Example\label{sec:Example}}

\begin{figure}[t!]
\centering \scalebox{0.8}{\includegraphics[scale=0.8]{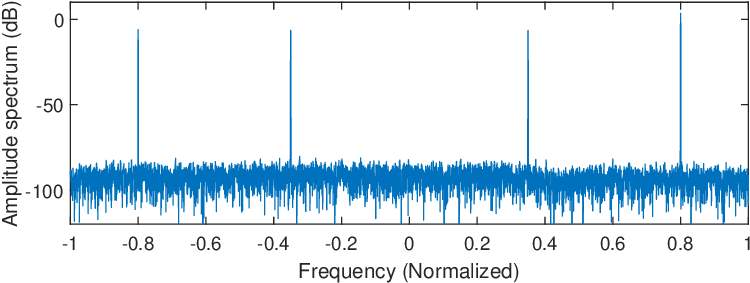}}
\caption{Spectrum of the reconstructed digital complex baseband signal.}

\label{fig:Design_example_spectrum}
\end{figure}

We consider the same scenario as in the first example of \cite{Wahab_2022},
viz. $M=2$, $d_{0}=0$ , and $d_{1}=-0.15$. The baseband signal
is a four-tone complex signal, according to (76) and (77) in \cite{Wahab_2022},
with additive noise corresponding to a signal-to-noise ratio (SNR)
of $61.8$ dB. Normalized with respect to $\omega T_{1}/\pi$, the
real bandpass signal has a bandwidth of $0.8$ centered at $5.15$,
thus covering parts of both the fifth and sixth Nyquist bands. Using
the design technique proposed in Section \ref{sec:Least-Squares-Design},
with $\omega_{1}T_{1}=4.75\pi$ and $\omega_{2}T_{1}=5.55\pi$, and
a filter order of $60$, the spectrum of the reconstructed signal
becomes as shown in Fig. \ref{fig:Design_example_spectrum}. The spurious-free
dynamic range (SFDR) is about $80$ dB whereas the SNR is some $59.6$
dB\footnote{Increasing the filter order beyond 60 offers a modest improvement
and the SNR saturates slightly below the original SNR of $61.8$ due
to a slight noise amplification through the reconstruction filters.}. For the technique in \cite{Wahab_2022} with filters of order $60$,
the SFDR and SNR are approximately $65$ and $58.2$, respectively\footnote{In \cite{Wahab_2022}, a mean-square error (MSE) of $-53.42$ was
reported. For the signal in the example, SNR $=-$MSE+4.77.}. For the proposal, those values can be reached with a filter order
of $46$. Hence, the proposed least-squares design is superior to
the windowing-based design in \cite{Wahab_2022}. 

\section{Conclusion\label{sec:Conclusion}}

This paper considered the problem of reconstructing digital complex
baseband signals via $M$-periodic nonuniform bandpass sampling and
time-varying FIR filters. A least-squares design method was proposed
and demonstrated to offer a lower reconstruction filter complexity
than that in an existing alternative reconstruction method. Ongoing
work encompasses a comprehensive study regarding filter order (complexity)
versus $M$, bandwidth, and reconstruction error, as well as finding
the optimal nonuniform sampling pattern when noise propagation is
taken into account. It also addresses the effect of undesired sampling
instance errors (time-skews), and more generally ADC-channel frequency
response mismatches, that will appear in practical nonuniform-sampling
TI-ADCs \cite{Wahab_2022}. When the ADC channel characteristics change,
the filters need to be modified (redesigned) in real time and it is
desired to minimize the complexity of the update. This requires efficient
reconfigurable reconstructors, which have been developed for regular
uniform-sampling TI-ADCs and real signals \cite{Johansson_07,Tertinek_08,Vogel_09,Johansson_09,Wang_2022}.
They are however not applicable here and it thus remains to develop
such reconstructors for complex baseband signals. The design technique
proposed in this paper serves as a starting point for this development.

\bibliographystyle{IEEEtran}
\bibliography{bibliography2}

\end{document}